\documentclass[12pt]{article}
\usepackage{latexsym,hyperref}
\usepackage{graphicx}
\usepackage{amssymb}
\usepackage{amsmath}
\usepackage{bm}
\usepackage{multirow}
\usepackage{subfigure}
\usepackage{wrapfig}
\usepackage{url}
\usepackage{enumitem}
\usepackage{relsize}
\usepackage[pdftex,dvipsnames]{xcolor}
\usepackage{longtable}
\usepackage{textcomp}

\usepackage[margin=1in]{geometry}
\usepackage{makecell}
\usepackage{multicol}
\usepackage{float}
\usepackage{lscape}
\usepackage{amsmath}
\usepackage[font=small,labelfont=bf]{caption}
\hypersetup{colorlinks=true,linkcolor=black}

\newcommand{\R}{\mathbb{R}}

\newcommand{\bfb}{{\bf b}}

\newcommand{\bfe}{{\bf e}}
\newcommand{\bff}{{\bf f}}

\newcommand{\bfm}{{\bf m}}

\newcommand{\bfr}{{\bf r}}
\newcommand{\bfs}{{\bf s}}

\newcommand{\bfw}{{\bf w}}
\newcommand{\bfx}{{\bf x}}

\newcommand{\bfC}{{\bf C}}

\newcommand{\bfE}{{\bf E}}
\newcommand{\bfF}{{\bf F}}

\newcommand{\bfK}{{\bf K}}

\newcommand{\bfM}{{\bf M}}

\newcommand{\bfX}{{\bf X}}

\newcommand{\bfW}{{\bf W}}

\newcommand{\beq}{\begin{equation}}
\newcommand{\eeq}{\end{equation}}
\newcommand{\beqs}{\begin{eqnarray}}
\newcommand{\eeqs}{\end{eqnarray}}

\newcommand{\calD}{{\cal D}}

\newcommand{\calG}{{\cal G}}

\newcommand{\calN}{{\cal N}}
\newcommand{\calP}{{\cal P}}

\DeclareMathOperator*{\argmax}{arg\,max}

\setcellgapes{10pt}
\begin{document}
\bibliographystyle{unsrt}
\begin{center}
\Huge
{\bf A machine-learning optimized vertical-axis wind turbine} \\

\vspace{5mm}
\normalsize

\vspace{2mm}
Huan Liu\footnote{Department of Mechanical and Civil Engineering, California Institute of Technology, Pasadena, CA 91125, USA} and Richard D. James\footnote{Department of Aerospace Engineering and Mechanics,
University of Minnesota, Minneapolis, MN 55455, USA

\;\;\;huanliu@caltech.edu,  james@umn.edu
} 
 \\

\end{center}

\vspace{5mm}
\noindent {\normalsize {\bf Abstract.} } Vertical-axis wind turbines (VAWTs) have garnered increasing attention in the field of renewable energy due to their unique advantages over traditional horizontal-axis wind turbines (HAWTs). However, traditional VAWTs including  Darrieus and Savonius types suffer from significant drawbacks -- negative torque regions exist during rotation. In this work, we propose a new design of VAWT, which combines design principles from both Darrieus and Savonius but addresses their inherent defects. The performance of the proposed VAWT is evaluated through numerical simulations and validated by experimental testing. The results demonstrate that its power output is approximately three times greater than that of traditional Savonius VAWTs of comparable size. The performance of the proposed VAWT is further optimized using machine learning techniques, including Gaussian process regression and neural networks, based on extensive supercomputer simulations. This optimization leads to a 30\% increase in power output.

\section{Introduction}
Wind carries significant amounts of energy and can serve as a mainstay of the future global energy system \cite{golston2019wind}. Wind turbines, rotating about either a horizontal or a vertical axis, are a medium that converts the kinetic energy of wind into electrical energy. 
Compared to the horizontal-axis wind turbine (HAWT), the vertical-axis wind turbine (VAWT) has yet to achieve widespread commercial success primarily because of the lack of good structural design and the complexity
of the underlying fluid dynamics \cite{miller2021solidity}. However, there is no evidence that the performance of HAWTs is inherently better than that of VAWTs. Unlike HAWTs, VAWTs are characterized by their rotational axis being perpendicular to the wind direction, which offers operational and design benefits. These include the ability to capture wind from any direction without the need for a yaw control system, and a lower center of gravity which can lead to enhanced stability, reduced structural demands, and simplified maintenance \cite{johari2018comparison}.

Historically, the development of VAWTs dates back to the early 20$^{th}$ century, with pioneering designs such as the Darrieus \cite{jin2015darrieus} and Savonius \cite{akwa2012review} rotors. The Darrieus rotor, known for its egg-beater shape, provides high efficiency under optimal high wind conditions; while the Savonius rotor, with its semi-cylindrical design, is often used for low-speed, low-torque applications. Advances in materials science and aerodynamics have significantly improved the performance and feasibility of VAWTs in recent years, leading to renewed interest in their potential for urban and offshore applications \cite{kumar2018critical}.

Recent studies have highlighted the advantages of VAWTs in urban environments, where their compact design and capabilities to capture wind from all directions make them suitable for integration into buildings and other structures. This is particularly relevant given the growing trend towards decentralized energy production and the need for innovative solutions to harness wind energy in built environments \cite{kumar2018critical}. Additionally, VAWTs have shown promise in offshore wind farms, where their lower height and reduced sensitivity to turbulent wind conditions can be beneficial \cite{borg2014offshore}. Despite these advantages, VAWTs face challenges related to efficiency and durability. Issues such as lower aerodynamic efficiency compared to HAWTs and the impact of cyclic loading on structural integrity need to be addressed through ongoing research and technological development \cite{johari2018comparison,saad2014comparison}. Therefore, exploring new generations of VAWT appears to be particularly important.

The main performance metric of wind turbines is power efficiency, which represents the ratio of mechanical energy in the wind turbine converted from the kinetic energy in the wind and is formulated by
\beq
\eta=\frac{P}{\frac{1}{2}\rho A V_w^3},\;\;{\rm with}\;\;P=\bfM\cdot\bfw,\label{efficiency}
\eeq
where $\bfM$ is the total torque of the turbine, $\bfw$ is the angular velocity of the rotor, $V_w$ is the free stream wind velocity, $\rho$ is the wind density, and $A$ is the swept area of the rotor. The strategy for improving efficiency is to increase the value of mechanical power $P$, which can be realized by increasing the effective torque or angular velocity. But the efficiency ratio will be bounded by Betz limit \cite{rauh1984betz}, which is $59.3\%$.

Dynamic similarity is a powerful tool to study the dynamic performance of large-scale structures in fluids. The scaling laws of interacting
fluids, elastic solids, and rigid bodies
are collected in the previous work \cite{liu2023options}.  Assume there is a laboratory model that is dynamically similar to the full-scale model, which involves three scaling parameters -- $\lambda_l$ for geometry, $\lambda_t$ for time, and $\lambda_\rho$ for density.  Let $P, \bfM$ and $\bfw$ denote the mechanical power, total torque, and angular velocity of the rotor for the laboratory wind turbine, and let $P^{(\lambda_l,\lambda_t,\lambda_\rho)},\bfM^{(\lambda_l,\lambda_t,\lambda_\rho)},\bfw^{(\lambda_t)}$ denote the corresponding quantities of the full-scale wind turbine, respectively. The scaling of torque and power are then found by
\beq
\begin{array}{lll}
\bfM^{(\lambda_l,\lambda_t,\lambda_\rho)}   &=&  \frac{\lambda_\rho\lambda_l^5}{\lambda_t^2}\bfM,\quad{\rm and}\\
  P^{(\lambda_l,\lambda_t,\lambda_\rho)}   &=&\bfM^{(\lambda_l,\lambda_t,\lambda_\rho)}\cdot\bfw^{(\lambda_t)}=\frac{\lambda_\rho\lambda_l^5}{\lambda_t^3}P.
\end{array}
\label{power_scaling}
\eeq
Since the power efficiency $\eta$ is a dimensionless number, its value for a full-scale model will be the same as that of the corresponding laboratory model, which can be verified by \cite{liu2023options}.

In this work, we propose a new VAWT design inspired by our recent work on curved tile origami \cite{liu2022origami,liu2024design,liu2025foldable}, which is composed of two rotors and one deflector. The performance of the VAWT is systematically studied by using computational simulations, which show better performance than traditional VAWTs. Its power output is then optimized by employing machine learning (ML) algorithms including Gaussian Process Regression and Neural Networks.

\section{The proposed vertical-axis wind turbine}
\begin{wrapfigure}{r}{0.5\textwidth}
\vspace{-8mm}
\begin{center}
\includegraphics[width=0.5\textwidth]{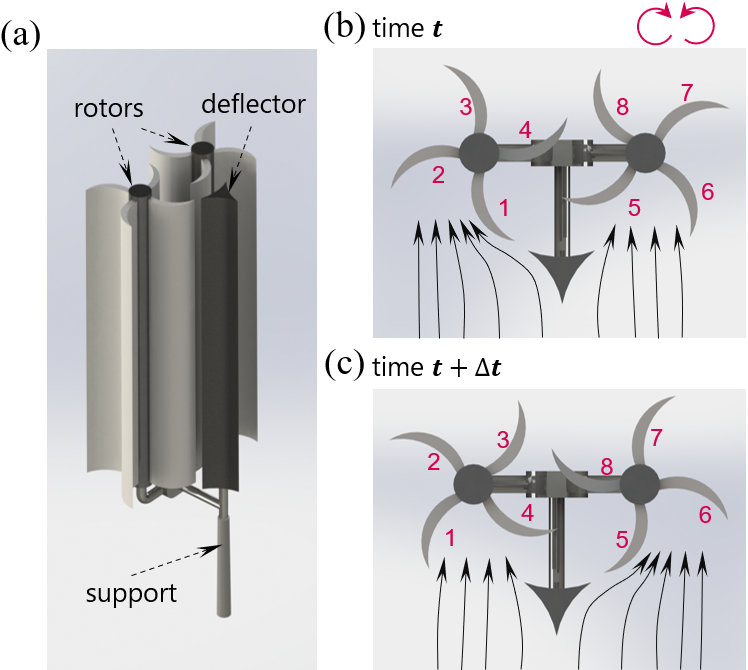}
\vspace{-1mm}
\caption{\small (a) Schematics of the proposed VAWT. The top views of the VAWT at different times $t$ (b) and $t+\Delta t$ (c), where the red arrows represent the sense of rotation of the two rotors and the black arrows indicate the wind flow.
\label{fig:top_view}}
\end{center}
\vspace{-8mm}
\end{wrapfigure}

In this section we present the new VAWT design. As the schematic shown in Figure \ref{fig:top_view}, the proposed VAWT \cite{vawtpatent} contains two meshed rotors, one smooth-faced deflector, one support at the base of the structure, and other necessary gearboxes and accessories (hidden). The deflector is mounted in the central front of the rotors to divert wind flow to the two rotors; the rotors and deflector are fitted with bearings into the support at their base. Together with the overall eccentricity of the structure, the assembly including the rotors and the deflector will rotate to align automatically with the wind direction as shown in Figure \ref{fig:top_view}(b) and (c) in which the black arrows represent the wind flow. For each rotor there are four blades, and the two curved surfaces in each blade are cylindrical. The two rotors have the same profile but reversed chirality, which will rotate in reverse directions in the wind, as indicated by the red arrows. The two rotors are transmitted via a gearbox to guarantee the same rotation rate. The rotations of rotors are then transmitted to an electric generator (hidden).  The support at the base will be embedded in poured concrete.

The proposed VAWT aims to improve energy efficiency by eliminating the negative torque in the parasitic regions of both rotors by using the deflector. 
Under certain wind conditions, the mechanism can be explained by Figure \ref{fig:top_view}(b) and (c) as follows. Figure \ref{fig:top_view}(b) shows the structure at a certain time $t$. At this time, Blade 1 is just
passing the deflector, which is acting as an airfoil, or equivalently, the
mainsail of a sailboat.  At the same time, Blade 2 is transitioning from airfoil to spinnaker, that is, from a structure that
generates torque from aerodynamic lift forces to a structure that generates torque from drag forces.
More precisely, the spinnaker mode generates torque by reducing the kinetic energy of the impinging air
from the free-stream value to a small value and converting this energy to mechanical energy with a power = torque exerted on the wind turbine $\times$ rotors' angular velocity. We
use the suggestive terminology of mainsail and spinnaker below for these two modes. Two of
the blades on the right rotor, 6 and 7, are similarly functioning as mainsail and spinnaker.
At a later time $t+\Delta t$ we refer to Figure \ref{fig:top_view}(c). At this time, Blade 5 of the other rotor
aligns with the deflector and functions as a mainsail, and Blade 6 is functioning as a spinnaker. Essentially,
these features imply that wind flow that would have been parasitic had the deflector been absent now contributes substantially to the torque. At the time $t+\Delta t$, Blade 1 is transitioning from a mainsail to a spinnaker.
Blade 8, which appears to be entering a quiescent region is in fact generating torque due to the meshed arrangement of the two rotors. Highly relevant to this aspect and other
cases above is flow separation. These features have been seen from the simulations that we will discuss later.


\section{Computational analysis}
\subsection{Basic setup}{\it Computational model.}
The computations are performed using the commercial CFD solver ANSYS Fluent 2020R2. The accuracy of the computational torque and power
used to evaluate the performance of the VAWT highly depends on the computational model. During the operation of the VAWT,  turbulence happens near the blades. We use the $SST$ (Shear Stress Transport) $k-\omega$ model \cite{menter1994two,menter1993zonal,menter2006correlation,langtry2006correlation,menter2009review}, which has proven to be particularly efficient in VAWT \cite{nobile2014unsteady}, to capture the turbulence.  This model combines two different models: $k-\omega$ for the inner boundary layer and $k-\epsilon$ for the free stream.

\begin{figure}[ht]
\begin{center}
\includegraphics[width=0.7\textwidth]{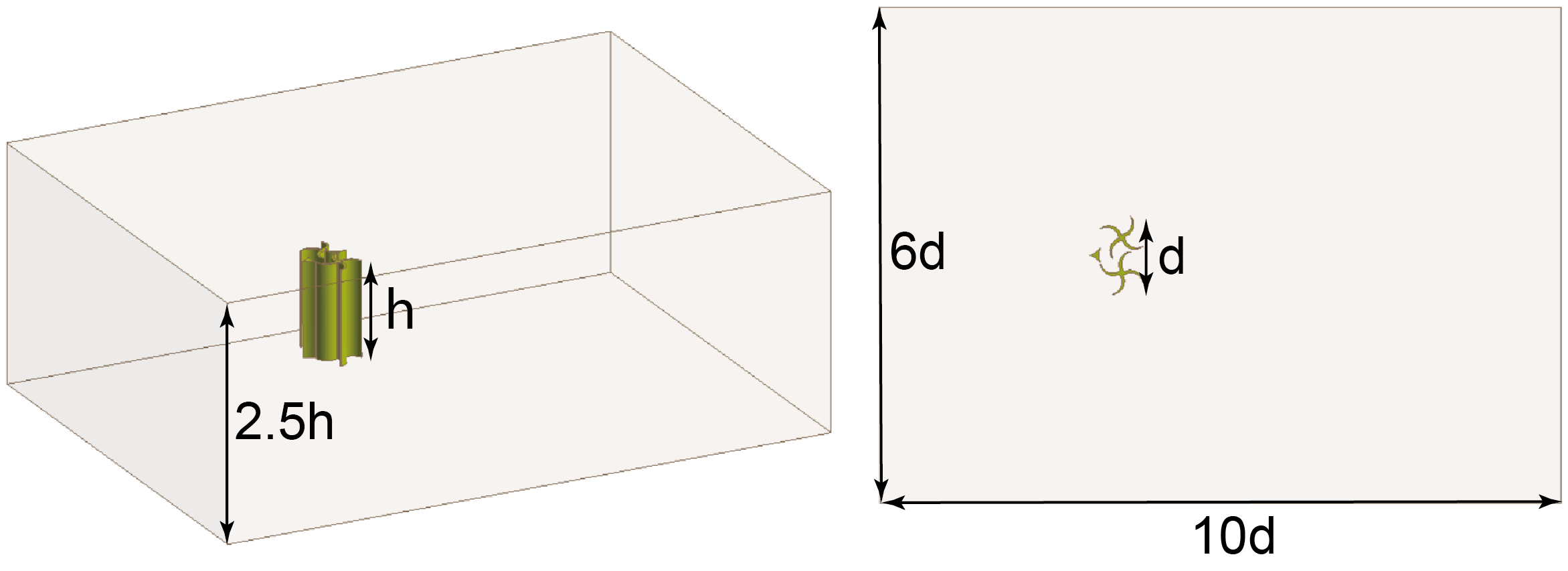} 
\caption{\small Schematics of the computational domain with dimensions, side view on the left and top view on the right.
}
\label{fig:domain}
\end{center}
\vspace{-5mm}
\end{figure}

{\it Computational domain.}
A 3D vertical-axis wind turbine model built in Figure \ref{fig:domain} is based on the design outlined in Figure \ref{fig:top_view}. The full-scale 3D model is used in the simulations by placing the VAWT within a dynamically meshing cubic region. To reduce the computational load, only the two rotors and the deflector are included in the model. Different domain dimensions
have been investigated in \cite{mclaren2011numerical} and the dimensions shown in Figure \ref{fig:domain} were determined to be sufficiently distant from the turbine to avoid the influence of the boundaries. Therefore, this model will be used for all computations in the present study.  For the full 3D  model, the upstream face is set as the velocity inlet boundary, the downstream face is set as the pressure outlet boundary, and the four side faces are set as no-slip walls. To study the power efficiency, we assign the angular velocity $\bfw$ and measure the total torque $\bfM$ acting on the rotors due to the wind. The total torque $\bfM$ along the rotation axis and power $P$ are found by
\beq
\bfM=\bfF_1\times\bfr_1-\bfF_2\times\bfr_2, \quad P=\bfM\times\bfw,
\eeq
where $\bfF_1$ and $\bfF_2$ are resultant forces acting on the two rotors, $\bfr_1$ and $\bfr_2$ are distances from the origins of $\bfF_1$ and $\bfF_2$ to the corresponding rotation axes, and the minus sign is caused by the counter-rotating fact. The power efficiency is then found by (\ref{efficiency}).

\begin{figure}[htp]
\begin{center}
\includegraphics[width=0.9\textwidth]{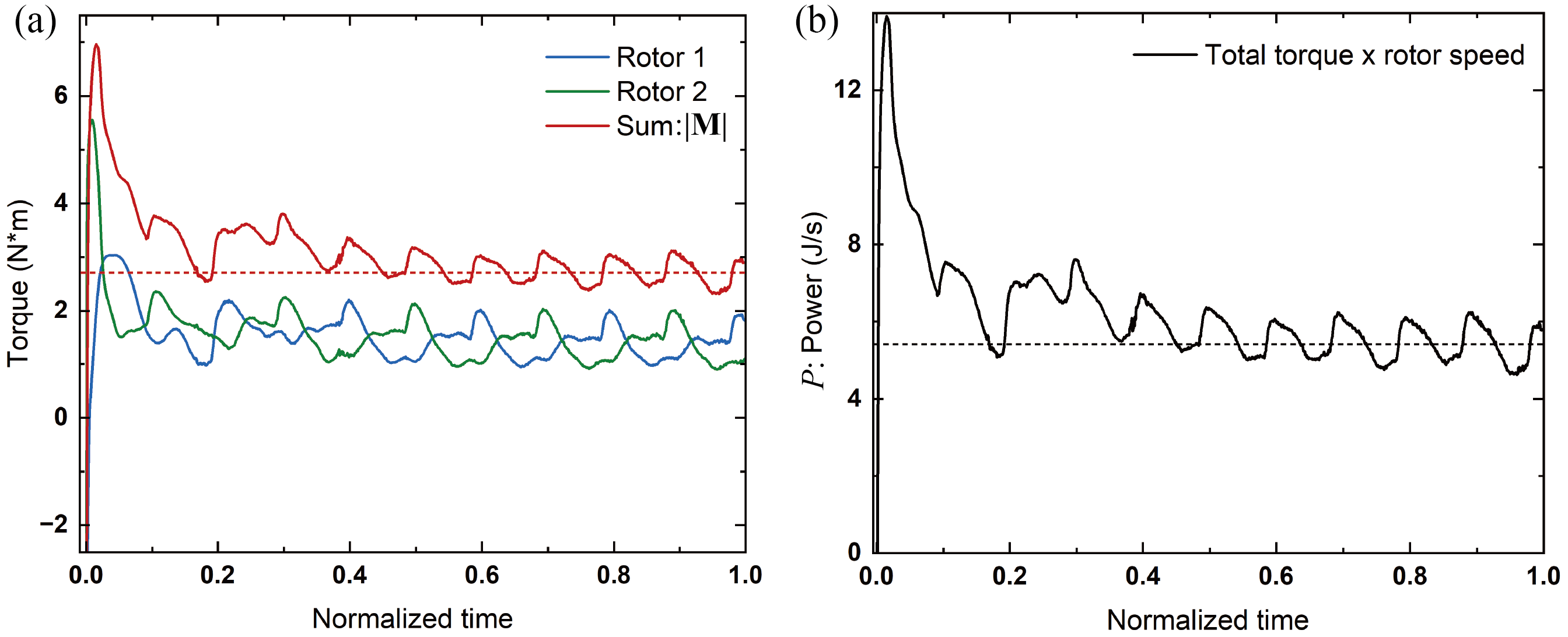} 
\caption{\small A typical example to show the torque (a) and power output (b) of the proposed VAWT during rotation. The dashed lines represent the mean values of total torque and power output respectively after the oscillatory flow is convergent.
}
\label{fig:example}
\end{center}
\vspace{-5mm}
\end{figure}

{\it Computational results.}
Transient simulations are conducted to examine the dynamic performance of the VAWT. The typical results, showing the torque and power output over time, are presented in Figure \ref{fig:example}. In this example, the wind speed is set to $3~m/s$, the rotor speed is set to $2~rad/s$, and the swept area of VAWT is chosen to  $1.5~m\times2~m$. From Figure \ref{fig:example}(a) the torques of the two rotors exhibit the same mean value, but a phase shift is present due to the asymmetric layout of the rotors. The total torque exerted on the VAWT is obtained by summing the two torques. The dashed red line represents the mean value of the total torque, as observed in the converged oscillatory flow. The power output is shown in Figure \ref{fig:example}(b), which is obtained by multiplying the total torque in (a) by the rotor speed $2~rad/s$, and the dashed black line shows the mean value of power output. In the following contents, we will use the mean values of total torque and power output as the primary indicators to characterize the performance of the VAWTs.

\begin{figure}[htp]
\begin{center}
\includegraphics[width=0.75\textwidth]{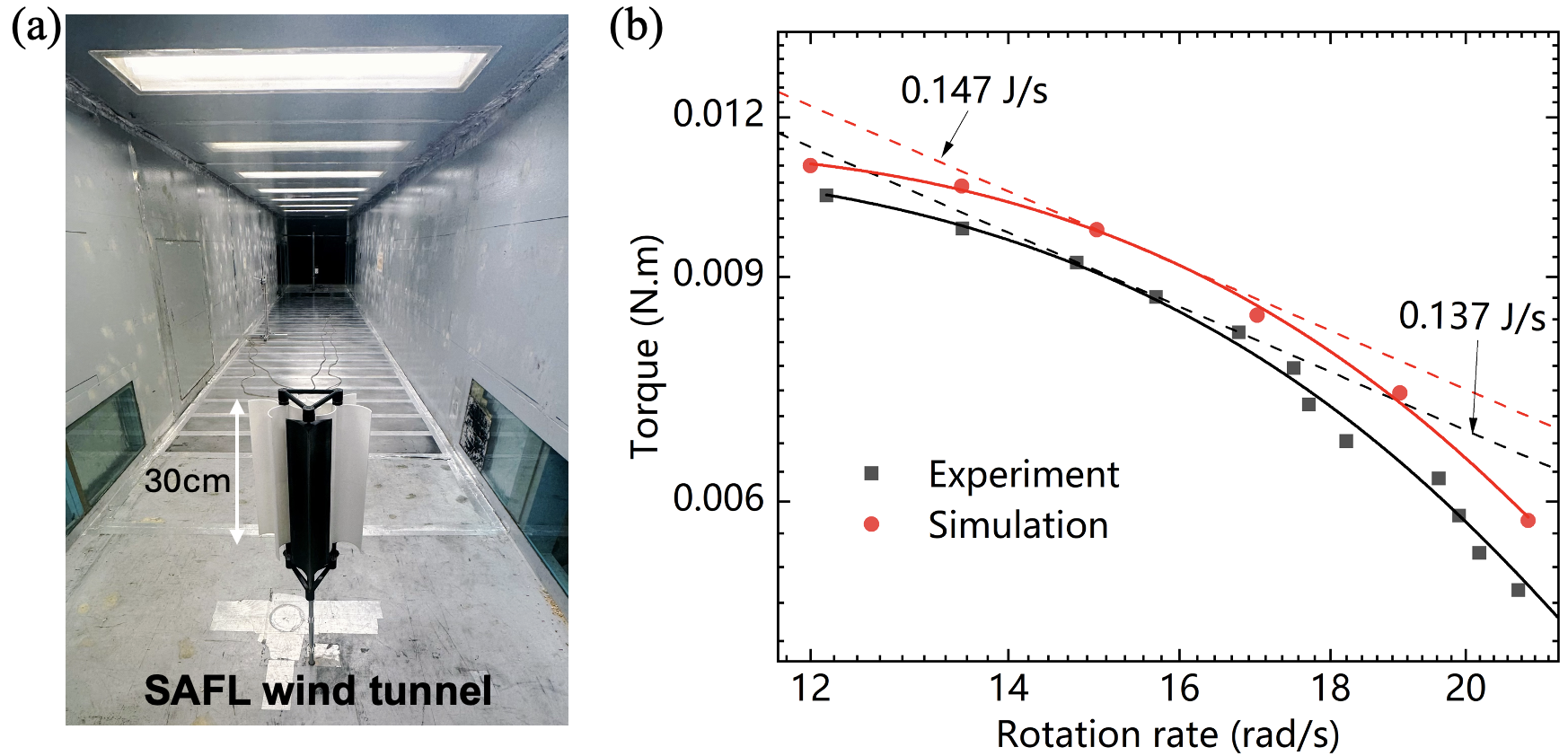} 
\caption{(a) Experimental setup of the proposed VAWT in the wind tunnel. (b) Comparison of simulated and experimental torque over time at a wind speed of $3~m/s$.}
\label{test}
\end{center}
\vspace{-5mm}
\end{figure}

{\it Experimental validation.}
To validate the computational model, we conducted wind tunnel tests at the St.~Anthony Falls Laboratory (SAFL). The experimental setup is depicted in Figure \ref{test}(a), where the prototype with a height of $30~cm$ was made of PLA using 3D printing technology and the wind tunnel dimensions are $1.5~m$ in height and $1.5~m$ in width. We regulate the (mean) torque applied to the turbine with a magnetic particle brake and measure the rotors' rotation rate after stabilization using a laser tachometer. 
Figure \ref{test}(b) presents the representative experimental and simulated results for a wind speed of $3~m/s$, which show good agreement, with the exception of a minor systematic error attributed to the accuracy of the gear fitting. The maximum power is 
then found by the tangent dashed lines, which represent constant power output and illustrate the inversely
proportional relationship between torque and rotation rate. The error in the maximum power is then estimated to be around 7\%, which is considered reasonable.

\subsection{Performance analysis}

In this subsection, we evaluate the performance of the proposed VAWT. First, we compare its performance with that of a traditional Savonius-type VAWT of similar size. Next, we systematically examine the power output of our design in relation to turbine size and wind speed. 

\begin{figure}[htp]
\begin{center}
\includegraphics[width=\textwidth]{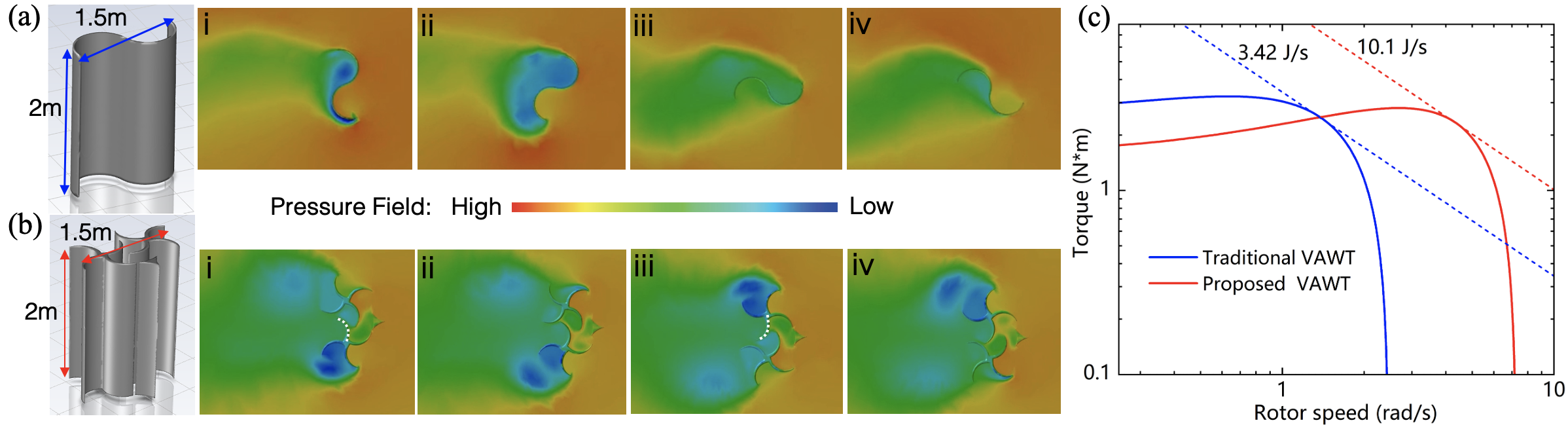} 
\caption{Snapshots of pressure fields of (a) the traditional VAWT and (b) the proposed VAWT during rotation, where the wind speed and the rotor speed are set to $3~m/s$ and $1~rad/s$.  (c) The torque-rotor speed plot of the two VAWTs.}
\label{fig:comparison}
\end{center}
\vspace{-5mm}
\end{figure}

{\it Comparison to traditional VAWT.} To better understand the fluid dynamics around the proposed and Savonius-type VAWTs, we first examine the pressure fields around them during rotation, which are shown in Figure \ref{fig:comparison}(a) and (b). The Savonius-type VAWT consists of two semi-cylindrical shells. Its surrounding pressure field shows that the blade on the parasitic side constantly generates negative torque, hindering a high power output. In contrast, the deflector and the meshed blades minimize the region generating negative torque on the parasitic side of the proposed VAWT with a comparable size. In particular, in Figure \ref{fig:comparison}(b) i and iii, the blades highlighted by the white dashed curves on the parasitic side generate positive torque. These effects enhance the performance of the VAWT, resulting in a rated (maximum) power output (10.1 J/s) that is nearly three times that of the traditional design (3.42 J/s), see Figure \ref{fig:comparison}(c).

{\it Performance of the proposed VAWT.} To systematically evaluate the performance of the proposed VAWT, we aim to determine its rated power output at a given wind speed and establish scaling laws that relate the rated power output to turbine size and wind speed. The following procedure is proposed to achieve this goal.
\begin{enumerate}[leftmargin=*]
    \item With the wind speed and the turbine size fixed, prescribe the rotation rate of the two rotors and perform the transient simulation to get the average torque output during rotation. Vary the rotor speed to get the relation between torque and rotor speed. Then find the rated power output at the given wind speed and turbine size.
    \item With the wind speed fixed, vary the turbine size. For each geometry size, repeat Step 1 to get the relation between the rated power output and the turbine size at a specific wind speed. 
    \item Vary the wind speed, and repeat Step 2 to get the scaling law of the rated power output with respect to the wind speed and geometry size.
\end{enumerate}
\begin{figure*}[htp]
\begin{center}
\includegraphics[width=0.92\textwidth]{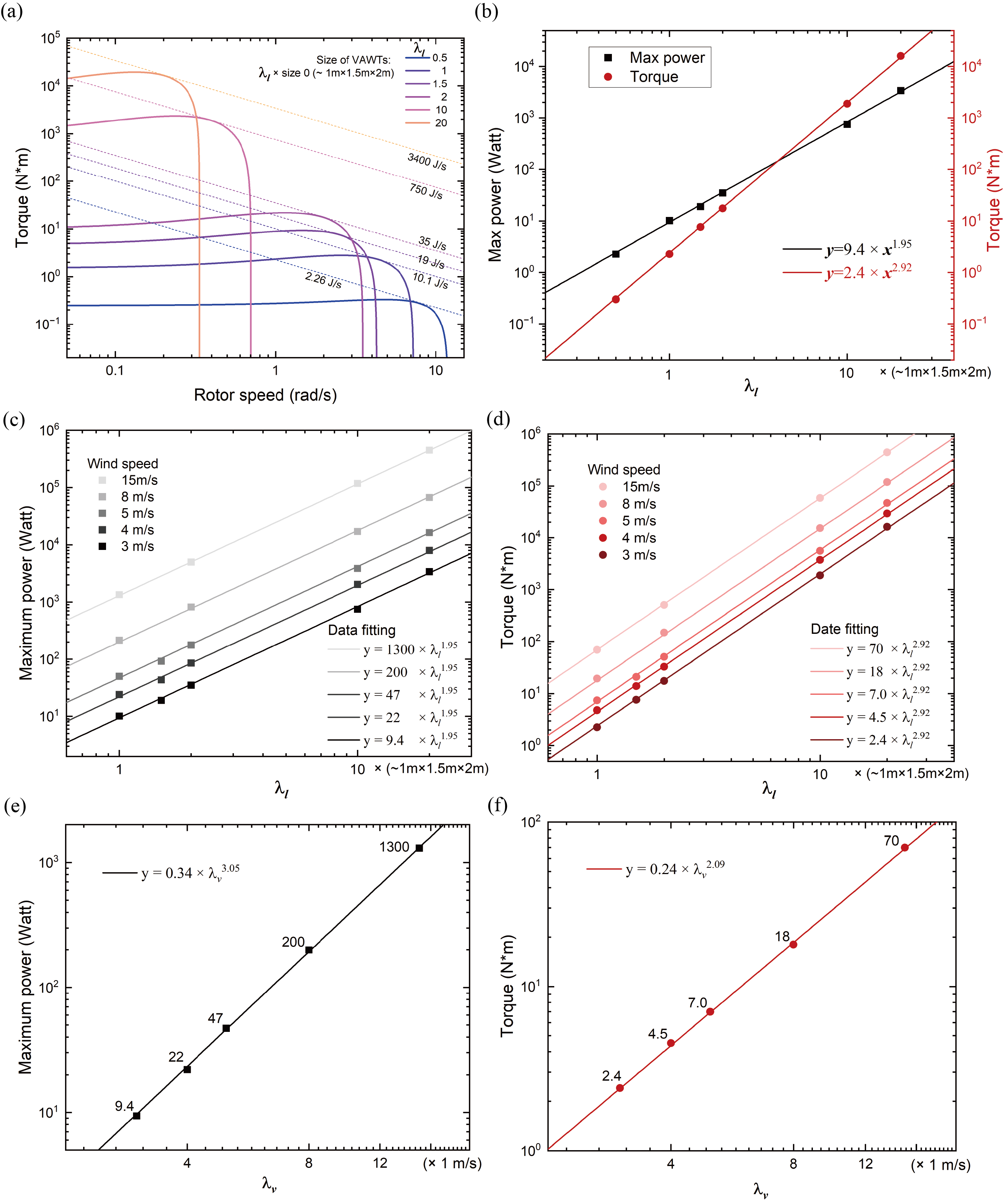} 
\caption{(a) The torque acted on VAWTs by the wind ($3m/s$) at different rotor speeds; Different curves correspond to different turbine sizes with size scaling parameter $\lambda_l$. (b) Data fitting of maximum (rated) power and corresponding rated torque from (a) with respect to the size of VAWTs. (c) The relation between maximum (rated) power and turbine sizes at wind speeds of 15 m/s,  8 m/s, 5 m/s, 4 m/s, and 3 m/s, respectively. (d) The relation between rated torque and turbine size at wind speeds of 15 m/s,  8 m/s, 5 m/s, 4 m/s, and 3 m/s, respectively. (e) The relation between maximum (rated) power and the wind speed at $\lambda_l=1$. (f) The relation between rated torque and the wind speed at $\lambda_l=1$.
}
\label{scaling_power}
\end{center}
\vspace{-5mm}
\end{figure*} 

The computational results, shown in Figure \ref{scaling_power}, illustrate these steps. Note that Figure \ref{scaling_power} uses a log-log plot for a better demonstration. In (a-b) the wind speed is fixed at $3~m/s$, while the turbine size varies from a height of  $1~m$ to $40~m$. Each solid curve in (a) exhibits the torque-rotor speed relationship at a fixed turbine size, obtained through Step 1. The rated power and corresponding torque for each turbine size are indicated by the dashed lines and the tangential points, as summarized by the points in (b). These points suggest a power law between rated power and turbine size at a fixed wind speed: $P\propto \lambda_l^{1.95}$. By varying the wind speed and repeating Step 2, the rated power and torque as a function of turbine size at different wind speeds are determined, see the curves in (c) and (d). Furthermore, from the results in (c-d), the rated power and torque as a function of wind speed can be obtained for a fixed turbine size, as illustrated in the example shown in (e) and (f).

According to Figure \ref{scaling_power}, power laws characterize the rated power and torque with respect to turbine size and wind speed, that is,
\beq
P^{(\lambda_l,\lambda_v)}\propto \lambda_l^{1.95}\lambda_v^{3.05},\quad|\bfM^{(\lambda_l,\lambda_v)}|\propto \lambda_l^{2.92}\lambda_v^{2.09},\label{simulation}
\eeq
where $\lambda_l$ and $\lambda_v$ are the scaling parameters of geometry size and wind speed, respectively. More specifically,  the rated power $P$ extracted by wind turbine is explicitly expressed by
\beq
P=0.34\ \lambda_l^{1.95}\lambda_v^{3.05}.\label{power-laws}
\eeq
The underlying mechanics of (\ref{power-laws}) can be revealed by the scaling laws in (\ref{power_scaling}). We use one VAWT as a laboratory model and scale it by a factor of $\lambda_l$ to create a full-scale model, and then test their dynamic performance at suitable wind speeds with the scaling parameter $\lambda_v$. If the two models are dynamically similar, according to the scaling of the density, velocity, and kinematic viscosity of the fluids, the scaling parameter $\lambda_\rho,\lambda_t$ satisfy
\beq
\lambda_\rho=1,\quad\frac{\lambda_l}{\lambda_t}=\lambda_v,\quad\frac{\lambda_l^2}{\lambda_t}=1,\label{delta_alpha}
\eeq
which are overdetermined by the three equations. Notice that the viscosity term has a negligible effect at high Reynolds numbers, we ignore the scaling of kinematic viscosity and consider the two models to be dynamically similar. Then $\lambda_\rho,\lambda_t$ are only restricted by the first two equations of (\ref{delta_alpha}).
Substituting these relations to the scaling of torque and power in (\ref{power_scaling}), we get
\beq
P^{(\lambda_l,\lambda_v)}= \lambda_l^2 \lambda_v^3P,\quad\bfM^{(\lambda_l,\lambda_v)}=\lambda_l^3\lambda_v^2\bfM,\label{scaling}
\eeq
which validates the scaling laws obtained in (\ref{simulation}).

\section{Geometry optimization}
\begin{figure*}[htp]
\begin{center}
\includegraphics[width=0.92\textwidth]{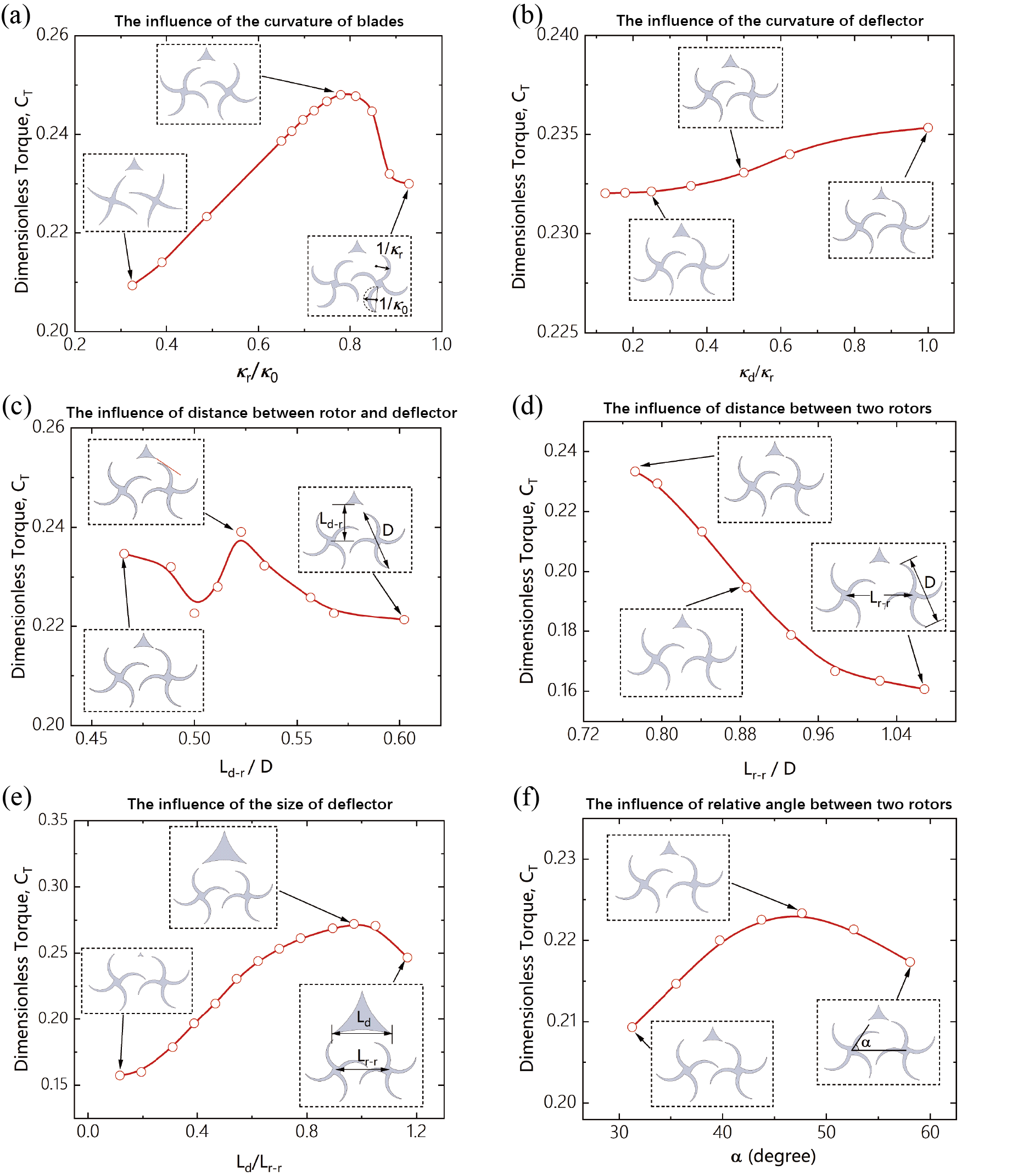} 
\caption{(a) The influence of the blade curvature $\kappa_r$ on the torque.  $1/\kappa_0$ is half of the length of the blades. (b) The influence of the deflector curvature $\kappa_d$ on the torque. (c) The influence of the distance $L_{d-r}$ between the rotor and deflector on the torque. $D$ represents the rotor diameter. (d) The influence of the distance $L_{r-r}$ between the two rotors on the torque. (e) The influence of the size $L_d$ of the deflector on the torque. (f) The influence of the relative angle $\alpha$ between the rotors on the torque.}
\label{fig:distance_rotors}
\end{center}
\vspace{-5mm}
\end{figure*}

In this section, we optimize the geometry of the proposed VAWT to maximize its power output. Here we fix the rotor height at $h=2m$, and set the rotor diameter to 0.44$h$. The geometry of the VAWT is then characterized by six main parameters: 1) the average curvature of the blades $\kappa_r$,  2) the curvature of the deflector $\kappa_d$, 3) the distance from the deflector to the plane defined by the axes of the two rotors  $L_{d-r}$, 4) the distance between the two rotors $L_{r-r}$, 5) the size of the deflector $L_d$, and 6) the relative angle between rotors $\alpha$. See notations in Figure \ref{fig:distance_rotors}. Let the rotor speed be set to 1 $rad/s$, and the wind speed be set to 3 $m/s$, where the Reynolds number is approximately $Re=8\times 10^4$. Let the design variable vector be
\beq
\bfx=(\kappa_r,\kappa_d,L_{d-r},L_{r-r},L_{d},\alpha).
\eeq
Let the characteristic distance of the two rotors be set to $0.34h$. Define the dimensionless torque coefficient by
\beq
C_T(\bfx):=\frac{M(\bfx)}{1/2\rho rA  V_w^2},
\eeq
where $r=0.44h/2$ is the radius of the rotor and $A=h(0.44h+0.34h)$. Let the design space be $\Omega_d$ as defined in Appendix B.
Our objective problem is to find a decision variable vector $\bfx^*\in\Omega_d$ such that $C_T(\bfx^*)$ is the maximum, that is,
\beq
\bfx^*=\argmax_{\bfx\in\Omega_d} \ C_T(\bfx)=\{\bfx:C_T(\bfx)\ge C_T(\bfs)\ {\rm for\ all\ } \bfs\in\Omega_d\}.\label{opti-vari}
\eeq
To visually see the effects of these parameters, we first independently study the influence of the six parameters on $C_T$ to identify the main parameters, and then use Gaussian Process Regression (GPR) and Neural Networks (NN) for
Machine Learning \cite{rasmussen2006gaussian,rasmussen2010gaussian} to study the coupling effects of the main parameters.

In Figure \ref{fig:distance_rotors}(a), $\kappa_0$ is constant representing the curvature of a circle with the length of the blade as its diameter. With the increase of the blade curvature, the torque first increases and then decreases. Figure \ref{fig:distance_rotors}(b) shows how the curvature of the deflector influences the torque, where $\kappa_d$ is the only variable representing the curvature of the deflector. As $\kappa_d$ increases, the total torque also increases but not significantly, and gets its maximum value as $\kappa_d\approx\kappa_r$. The influence of the distance between the rotor and the deflector is shown in Figure \ref{fig:distance_rotors}(c), where $D$ represents the diameter of the rotor. With the increase of $L_{d-r}$, the total torque presents first a decrease, then an increase, and then a decrease.  Figure \ref{fig:distance_rotors}(d) illustrates the influence of the distance $L_{r-r}$ between two rotors. As $L_{r-r}$ is small, the flow fields around the two rotors will interact with each other; as $L_{r-r}$ is great enough, the interaction is weak and the two rotors rotate independently. Figure \ref{fig:distance_rotors}(d) shows that the total torque decreases dramatically compared to Figure \ref{fig:distance_rotors}(a-b) with the increase of $L_{r-r}$. This implies the interaction between the two rotors improves the overall performance of the VAWT significantly. $L_d$ denotes the distance between the two corners of the deflector as shown in Figure \ref{fig:distance_rotors}(e). One can see the size of the deflector dramatically affects the torque and the torque gets its maximum value as $L_d\approx L_{r-r}$.  Figure \ref{fig:distance_rotors}(f) states the influence of the mismatch angle $\alpha$ between two rotors on the total torque. When $\alpha=0$, the two rotors are symmetric about the central plane. One can see that the total torque reaches its maximum when $\alpha$ is about $45^\circ$.

In summary, Figure \ref{fig:distance_rotors}(a-e) show that the curvature of blades $\kappa_r$, the distance between two rotors $L_{r-r}$, and the size of the deflector $L_d$ play major roles in performance of the proposed VAWT. However, the relation between the maximum $C_T$ and the parameter $L_{d-r}$ is unclear, and the influence of the combination of these parameters on the power output is also unclear.  Below, we use machine learning approaches to study the coupling effects of the six parameters and predict the optimal configuration of the VAWT.

\subsection{Machine-learning (ML) based optimization model}
\label{ML_models}
Our goal here is to find $\bfx^*$ and $C_T(\bfx^*)$ defined by (\ref{opti-vari}) via machine learning approaches on the design space $\Omega_d$. Among already established data-driven models suggested in the field of mechanics, we employ both Gaussian process regression (GPR) and neural network (NN) models to build predictive models. The training dataset $\calD=\{(\bfx_i,C_T(\bfx_i)):i=1,2,\dots,250\}$ contains 250 observations, informed by supercomputer  simulations. The GPR model is a non-parametric probabilistic supervised machine learning framework based on Bayesian inference. Despite the limited knowledge of the function $f(\bfx)$ between the input and output, a Gaussian process from some {\it a priori} belief is assumed on the underlying function. That is, 
\beq
f\sim\calG\calP(m,K),
\eeq
where $m$ is the mean function and $K$ is the covariance (kernel) function. For numerical implementation, it means for any arbitrary set of inputs, $\{\bfx_i,i=1,2,\dots,n\}$, the values of $f(\bfx)$ at these points, $\bff=[f(\bfx_1),f(\bfx_2),\dots,f(\bfx_n)]$, are correlated via a multivariate Gaussian distribution, i.e.,
\beq
\bff\sim \calN(\bfm,\bfK)
\eeq
where $\bfm=[m(\bfx_i)]_{n\times1}$ and $\bfK=[K(\bfx_i,\bfx_j)]_{n\times n}$.
According to Bayesian inference, a refined posterior distribution of $f$ can be derived by taking the observations into account. Since these observations in $\calD$ are inevitably contaminated with some noise when they are collected from simulations, a white noise, $\varepsilon\sim\calN(0,\sigma_0^2)$, is employed in the analysis, i.e., $C_T(\bfx_i)=f(\bfx_i)+\varepsilon(\bfx_i)$. Denote $(\bfX, \bfC_T)=(\bfx_i,C_T(\bfx_i))_{i=1,2\dots,n}$. According to the additivity of the Gaussian process, we have
\beq
C_T\sim\calG\calP(m,K_c), \quad K_c=K+K_\varepsilon,
\eeq
and
\beq
\left[\begin{array}{c}
     \bfC_T  \\
    C_T(\bfx^*) 
\end{array}\right]\sim\calN\left(\left[\begin{array}{c}
     \bfm(\bfX)  \\
    \bfm(\bfx^*) 
\end{array}\right],\left[\begin{array}{cc}
     \bfK_c(\bfX,\bfX)&\bfK_c(\bfX,\bfx^*)  \\
    \bfK_c(\bfx^*,\bfX)& \bfK_c(\bfx^*,\bfx^*)
\end{array}\right]\right),\label{gpr}
\eeq
where $K_\varepsilon(\bfx_i,\bfx_j)=\sigma_0^2\delta_{ij}$ is the noise kernel. Then from (\ref{gpr}) $C_T(\bfx^*)$ given $\bfC_T$ satisfies
\beq
C_T(\bfx^*)|\bfC_T\sim\calN(\bfm^*,\bfK^*),
\eeq
where the mean value is $\bfm^*=\bfK_c(\bfx^*,\bfX)\bfK_c^{-1}(\bfX,\bfX)(\bfC_T-\bfm(\bfX))+\bfm(\bfx^*)$ and the covariance function is $\bfK^*=\bfK_c(\bfx^*,\bfx^*)-\bfK_c(\bfx^*,\bfX)\bfK_c^{-1}(\bfX,\bfX)\bfK_c(\bfX,\bfx^*)$. The key to making predictions with a GPR model is selecting appropriate hyperparameters (parameters of a prior distribution), including the mean function $m(\bfx)$, the kernel function $K(\bfx,\bfx')$, and the noise $\sigma_0^2$. We set the noise term to nearly zero, say $\sigma_0=10^{-3}$, as the simulations have high fidelity and the simulation results of observations are kept at three decimal places. The mean function is assumed constant, i.e., $m(\bfx)=m_0$. We set $m_0$ to about 0.16 since it is the middle value of the observations. Then a widely used scaled squared exponential kernel model, $K(\bfx,\bfx')=\sigma^2 \bfe^{-\frac{|\bfx-\bfx'|^2}{2l^2}}$, is adopted. After optimizing hyperparameters by minimizing the negative marginal likelihood, a GPR model with $l=-0.5$ and $\sigma=0.5$ is proposed.

\begin{figure}[htp]
\begin{center}
\includegraphics[width=0.9\textwidth]{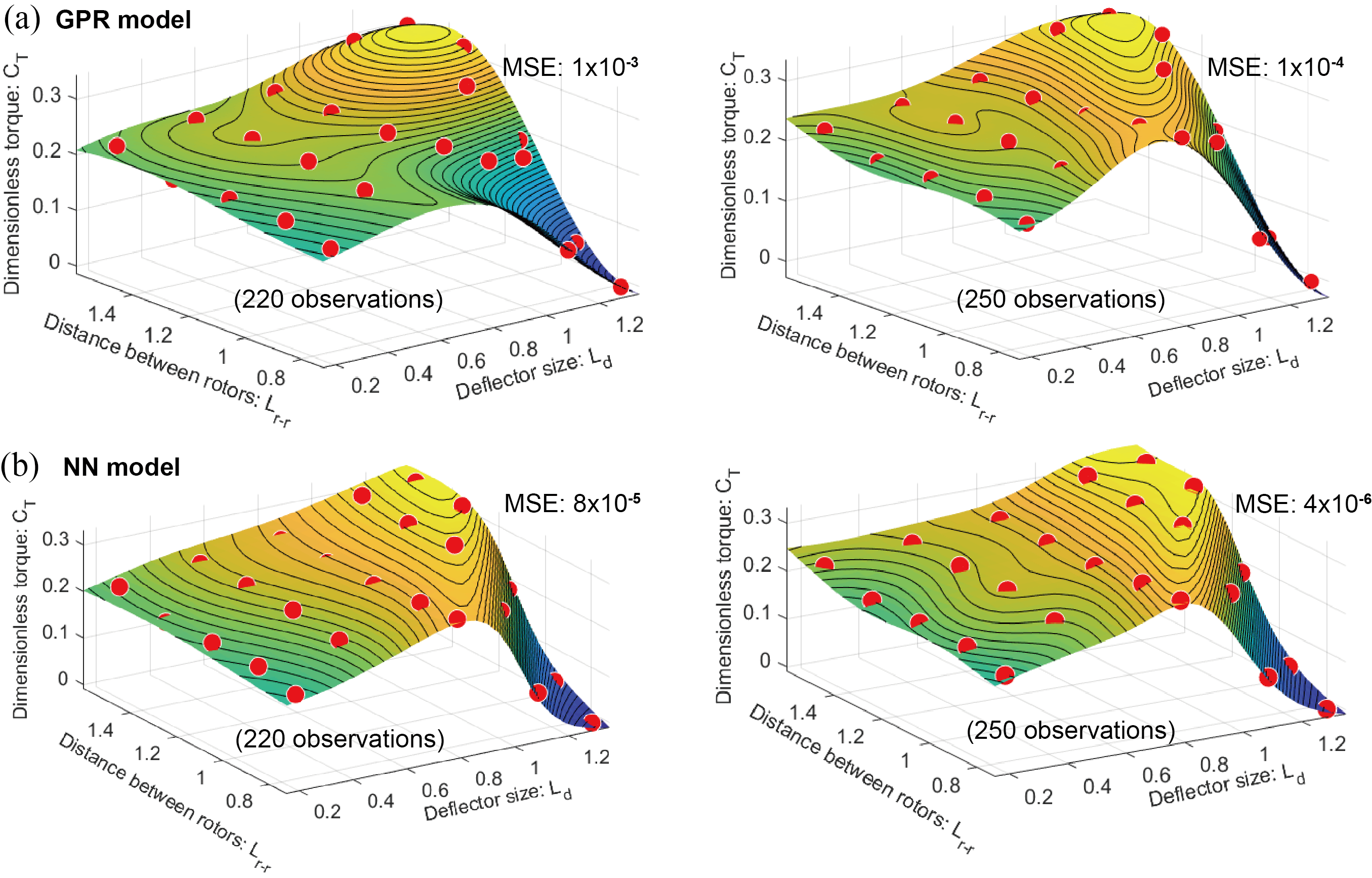} 
\caption{Contour plots predicted by the proposed Gaussian process regression (GPR) model and the neural network (NN) model, where the red dots are simulated data for testing.
\label{fig:gpr_ct}}
\end{center}
\vspace{-5mm}
\end{figure}

To evaluate the GPR model, we test it by using 220 observations and 250 observations from the training dataset, respectively. Figure \ref{fig:gpr_ct}(a) shows two predicted contour plots based on the two datasets, where $(L_{r-r},L_d)$ vary between $0.36\leq L_{r-r}\leq1.6,\ 0.15\leq L_d\leq1.35$ and the other four parameters $(\kappa_r,\kappa_d,L_{d_r},\alpha)$ are fixed at $(4.95,0.765,0.47,45^\circ)$. 24 points for testing are then simulated, which are highlighted in red in Figure \ref{fig:gpr_ct}. With 220 observations, the mean squared error (MSE) of the testing data is around $1\times 10^{-3}$ and the model shows large deviations at some testing points. As the number of observations increases to 250, MSE decreases to $1\times 10^{-4}$, which gives more accurate predictions.

To improve the reliability of the optimal results given by the GPR model, neural network-based (NN) deep-learning algorithms are also employed to study the latent function $f(\bfx)$. The concept of the neural network method is borrowed from neuroscience, which is composed of neural layers and neurons within each layer. The basic idea of NN is to approximate the complex latent function $f(\bfx)$ by the composition of a family of simple, continuous, piecewise affine (or linear) functions,  i.e.,
\beq
f(\bfx)=f_n\circ f_{n-1}\circ\dots\circ f_2\circ f_1 (\bfx),
\eeq
where $f_i,i=1,\dots,n$ is an affine function, $n$ is the number of the affine functions, and $f_1,\dots,f_{n-1}$ between the input $\bfx$ and output $f(\bfx)$ are so-called hidden layers. In the implementation, $f(\bfx)$ often has the form of 
\beq
f(\bfx)=f_n\circ\sigma\circ  f_{n-1}\circ\dots\circ\sigma\circ  f_2\circ \sigma\circ  f_1 (\bfx),
\eeq
with
\beq
\begin{array}{ll}
&   f_i=(\bfW_i|\bfb_i), \; \bfW_{i}\in\R^{p\times q},\;  \bfb_i\in\R^p,\\
&    p={\rm dim}(f_{i}(\cdot)),\;q={\rm dim}(f_{i-1}(\cdot)), 
\end{array}
\eeq
where $\sigma$ acts as a normalizer to rescale the output components of each layer (i.e., to get the scaled input of the next layer) and is called an actuation function, the matrix $\bfW_i$ is often called weights, $p$ is the dimension of $f_i(\cdot)$ and is also called the width of $i$th layer, and $q$ is the width of the previous layer. Typically, $\sigma$ is chosen as a sigmoid function, and $\sigma(x)=\frac{1}{1+e^{(-x)}}$ and $\sigma(x)=\tanh(x)$ are widely used ones, which squash values to the ranges of $(0,1)$ and $(-1,1)$, respectively. The role of $\sigma$ is to speed learning and also introduce nonlinearity to the system. For the NN method, the key is to get $\bfW_i$ and $\bfb_i, i=1,\dots,n$ that best describes $f(\bfx)$ by learning the training data. This is done by employing a gradient descent algorithm to minimize the mean squared error, say $\bfE(C_T,f(\bfx))$, between the predictions and the training set, where $C_T$ is the output from the training data. For instance, with a typical gradient descent algorithm, the weight $\bfW_i$ is iteratively updated by 
\beq
\bfW_i(t+1)=\bfW_i(t)-l_r\frac{\partial\bfE^t}{\partial\bfW_i},
\eeq
where $t$ represents an iteration step, and $l_r$ is the step size (also referred to as the learning rate). And $\frac{\partial\bfE^t}{\partial\bfW_i}$ is found by
\beq
\frac{\partial\bfE^t}{\partial\bfW_i}=\bfX_{i-1}\circ\sigma'\circ\bfW^{\rm T}_{i+1}\circ\sigma'\circ\dots\circ\bfW^{\rm T}_{n-1}\circ\sigma'\circ\bfW^{\rm T}_{n}\circ\sigma'\circ\frac{\partial\bfE^t}{\partial \bfX_n},\label{backprop}
\eeq
where $\bfX_i$ is the input of $i$th layer. The process described by (\ref{backprop}) is known as backpropagation.

Based on the same training datasets, we propose 32 models where the number of hidden layers $n$, the width of each layer $w_l$, and the learning rate $l_r$  of the model optimizer are defined by
\beq
\begin{array}{cc}
n\in \{1,2\},\; l_r\in\{0.01,0.001\}, \\
 w_l\in\{10,20,30,40,50,64,80,128\}
\end{array}
\eeq
and the $tanh$ function is selected as the actuation function for each layer. For the smaller dataset, the model with hyperparameter $(n,w_l,l_r)=(1,64,0.01)$ is selected by minimizing MSE for both training data and testing data. Similarly, for the datasets with 250 training points, the model with $(n,w_l,l_r)=(2,128,0.001)$ is then trained. Figure \ref{fig:gpr_ct} shows the contour plots predicted by the two NN models, with their respect MSE for the 24 testing points of $8\times10^{-5}$ and  $4\times10^{-6}$, which are much smaller than corresponding MSE from the GPR models. Thus, the NN model is better suited for this multivariable system. From the NN model, the $C_T$ can reach 0.336 within the preset design space, representing an approximately 30\% increase compared to the initial design.

Compared to traditional VAWTs, the proposed design demonstrates significant potential for practical application. By expanding the dimensions of geometric optimization, refining its topology, and integrating machine learning techniques, its performance is expected to be further enhanced.

\section*{Acknowledgment}The authors
acknowledge the support of the ONR MURI Grant FA9550-16-1-0566
and a Vannevar Bush Faculty Fellowship (N00014-19-1-2623). This work also benefited from AFOSR (FA9550-23-1-0093),
an ONR BIMADS project (N00014-23-1-2754), as well as a UMN MNBridge Grant. We thank Georgios Grekas for the discussion of the Neural Network algorithms, and the staff of the 
Venture Center at the University of Minnesota for their
support and advice.  We also thank the staff of the St. Anthony Falls Laboratory at the University of Minnesota for
providing wind tunnel facilities and technical help.

\appendix   

\section{Check of convergence}
\begin{figure}[htp]
\begin{center}
\includegraphics[width=\textwidth]{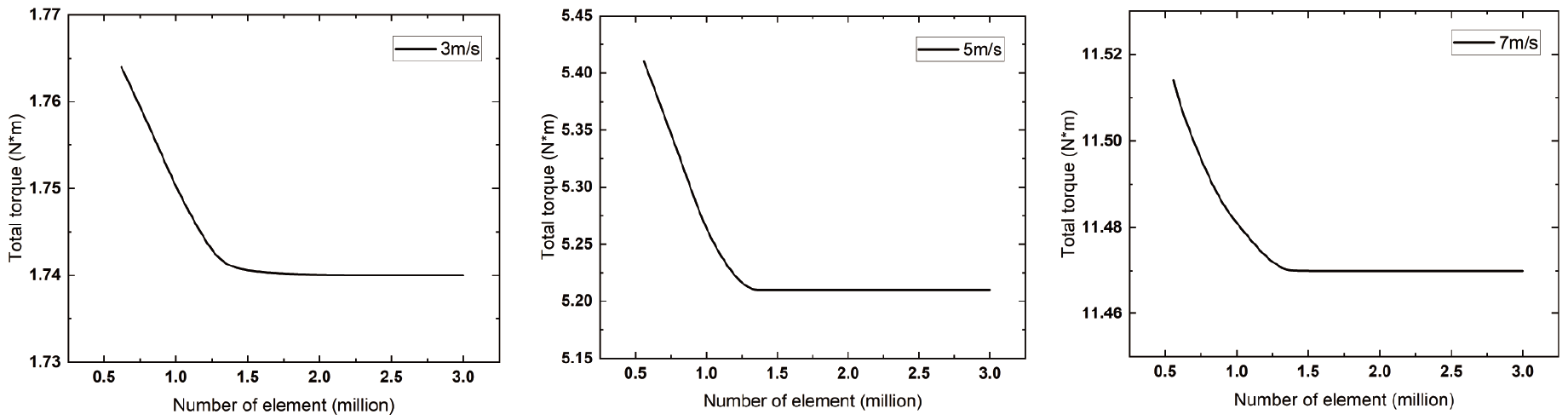} 
\caption{The convergence curves of total torque exerted on VAWTs in terms of the number of tetrahedron elements.}
\label{fig:mesh_number}
\end{center}
\vspace{-5mm}
\end{figure}

In the above simulations, we use tetrahedrons to discretize the simulation domain, and the number of tetrahedron elements varies between $1.37$ million and $2$ million. To validate the simulation results, the convergence of the variable of interest, i.e., the total torque exerted on the VAWT under prescribed rotor speed, in terms of the number of elements is studied and shown in Figure \ref{fig:mesh_number}. In the Figure, the rotor speed is set as $1rad/s$ and the wind speed is first set as $3m/s$, see the subfigure on the left. We find that 1) with the increase in the number of elements, the total torque is gradually convergent; and 2) the difference between the total torque and its convergent value is less than $1\text{\textperthousand}$ as the number of elements is greater than $1.3$ million. We also study the convergence curves of torque at wind speeds of $5m/s$ and $7m/s$, and we get the same conclusions, see the two right subfigures. Thus, as the number of elements increases to $1.37-2$ million, the tolerance of the total torque is acceptable.

\section{Design space}
The diameter of the rotor $D$ is fixed at $0.44 h$. The design space $\Omega_d$ of interest for the six parameters $\kappa_r$, $L_{r-r}$, $L_d$, $L_{d-r}$, $\alpha$, and $\kappa_d$ are defined by
\beq
\begin{array}{ll}
1\leq\kappa_r\leq 5.26, &0.7\leq L_{r-r}\leq1.5,\\0.1\leq L_{d}\leq1.3,&0.2\leq L_{d-r}\leq1,\\
0.1\leq\kappa_d L_d<1,&0<\alpha\leq45^{\circ}.
\end{array}\label{range1}
\eeq
To prevent the two rotors from colliding with the deflector and with each other during rotation, the six parameters are further subjected to
\beq
\begin{array}{ccc}
&\frac{1}{4}(L_{r-r}-L_d)^2+L_{d-r}^2>0.44^2,& \\
&L_{r-r}\cos(45^\circ-\frac{\alpha}{2})-\left(\frac{1}{\kappa_r}-((\frac{1}{\kappa_r})^2-0.19^2)^{\frac{1}{2}}\right)>0.44.&
\end{array}\label{range2}
\eeq

\bibliography{vawt.bib}
\end{document}